\documentstyle[aps,prl,psfig,epsfig,floats]{revtex}
\begin{document}
\draft
\twocolumn[\hsize\textwidth\columnwidth\hsize
\csname @twocolumnfalse\endcsname
\title{Electronic theory for the normal state spin dynamics in 
Sr$_2$RuO$_4$:\\
 anisotropy due to spin-orbit coupling}
\author{I. Eremin$^{1,2}$, D. Manske$^1$, and K.H. Bennemann$^1$}
\address{$^1$Institut f\"ur Theoretische Physik, Freie Universit\"at 
Berlin, D-14195 Berlin, Germany}
\address{$^2$Physics Department, Kazan State University, 420008 Kazan, Russia}
\date{\today}
\maketitle
\begin{abstract}
Using a three-band Hubbard Hamiltonian we calculate within the 
random-phase-approximation the spin susceptibility, $\chi({\bf q},\omega)$, 
and NMR
spin-lattice relaxation rate, 1/T$_1$, in the normal state 
of the 
triplet superconductor Sr$_2$RuO$_4$ and obtain quantitative agreement with
experimental data. Most importantly, we find that due to spin-orbit coupling
the out-of-plane component of the spin susceptibility $\chi^{zz}$ 
becomes at low temperatures two times larger than the in-plane one. As a 
consequence strong incommensurate antiferromagnetic fluctuations of the 
quasi-one-dimensional $xz$- and $yz$-bands point into the $z$-direction. 
Our results provide further evidence for the importance of 
spin fluctuations for triplet superconductivity in Sr$_2$RuO$_4$.
\end{abstract}
\pacs{74.20.Mn, 74.25.-q, 74.25.Ha}
]

\narrowtext
The spin-triplet superconductivity with $T_c$=1.5K observed  
in layered Sr$_2$RuO$_4$ seems to be a new example of unconventional 
superconductivity\cite{maeno}. The non $s$-wave symmetry of the order 
parameter is observed in several experiments (see for example 
\cite{kitaoka,duffy}). Although the structure of Sr$_2$RuO$_4$ is the same 
as for the high-T$_c$ superconductor La$_{2-x}$Sr$_x$CuO$_4$, its 
superconducting properties resemble those of superfluid $^3$He. 
Most recently it was found that the
superconducting order parameter is of $p$-wave type, but 
contains line nodes half-way 
between the RuO$_2$-planes\cite{tanatar1,izawa}. 
These results support Cooper-pairing via spin fluctuations as one of
the most probable mechanism to explain the triplet superconductivity in 
Sr$_2$RuO$_4$. Therefore, theoretical and experimental 
investigations of the spin dynamics behavior in the normal 
and superconducting state of Sr$_2$RuO$_4$ are needed. 

Recent studies by means of inelastic neutron scattering(INS)\cite{sidis} 
and nuclear magnetic resonance(NMR)\cite{mukuda} of the spin dynamics 
in Sr$_2$RuO$_4$ reveal the presence
of strong 
incommensurate fluctuations in the RuO$_2$-plane 
at the antiferromagnetic wave vector {\bf Q}$_i = (2\pi/3, 2\pi/3)$. 
As it was found in 
band structure calculations\cite{mazin}, they result from the 
nesting properties of the quasi-one-dimensional $d_{xz}$- and $d_{yz}$-bands. 
The two-dimensional $d_{xy}$-band contains only weak ferromagnetic 
fluctuations. The
observation of the line nodes between the RuO$_2$-planes\cite{tanatar1,izawa} 
suggests strong spin fluctuations 
between the RuO$_2$-planes in $z$-direction\cite{rice,annett,eremin}. 
However, inelastic neutron
scattering\cite{servant} observes that magnetic 
fluctuations are purely two-dimensional and originate from 
the RuO$_2$-plane. Both behaviors could result as a consequence 
of the magnetic anisotropy within the RuO$_2$-plane as indeed was observed 
in recent NMR experiments by Ishida {\it et al.} 
\onlinecite{ishida}. In particular, analyzing
the temperature dependence of the nuclear 
spin-lattice relaxation rate on $^{17}$O in the RuO$_2$-plane at low 
temperatures, they have demonstrated that the out-of-plane 
component of the spin susceptibility can become 
almost three time larger than the in-plane one. 
This strong and unexpected anisotropy disappears with increasing
temperature\cite{ishida}. 

In this Communication we analyze the normal state spin dynamics of the
Sr$_2$RuO$_4$ using the two-dimensional 
three-band Hubbard Hamiltonian for the three bands crossing the Fermi
level. We calculate  the dynamical spin 
susceptibility $\chi({\bf q}, \omega)$ within the random-phase-approximation 
the and show that the observed magnetic anisotropy in the
RuO$_2$-plane arises mainly due to the spin-orbit coupling. Its further
enhancement with lowering temperatures is due to the vicinity to a magnetic 
instability. Thus, we demonstrate that as in the superconducting
state\cite{ng} the spin-orbit coupling plays an important role also for the
normal state spin dynamics of Sr$_2$RuO$_4$. We also discuss briefly 
the consequences of this magnetic anisotropy for Cooper-pairing 
due to the exchange of spin fluctuations. 

We start from the two-dimensional three-band Hubbard Hamiltonian
\begin{equation}
H=H_{t}+H_{U}=\sum_{{\bf k}, \sigma} \sum_{l} t_{{\bf k} l} 
a_{{\bf k}, l \sigma}^{+} a_{{\bf k}, l \sigma} +
 \sum_{i,l} U_{l} n_{i l \uparrow} n_{i l \downarrow},
\label{hamilt}
\end{equation}
where $a_{{\bf k}, l \sigma}$ is the Fourier-transformed 
annihilation operator for the $d_{l}$ orbital electrons
($l = xy, yz, zx$) and $U_l$ is the corresponding 
on-site Coulomb repulsion. $t_{{\bf k} l}$ denotes the energy 
dispersions of the tight-bindings bands calculated as follows:
$t_{{\bf k} l} = - \epsilon_0 -2t_x \cos k_x - 2t_y \cos k_y
+4t' \cos k_x \cos k_y$. We choose the 
values for the parameter set ($\epsilon_0, t_x, t_y, t'$) as 
(0.5, 0.42, 0.44, 0.14), (0.24, 0.31, 0.045, 0.01), and 
(0.24, 0.045, 0.35, 0.01)eV for $d_{xy}$-, $d_{zx}$-, and $d_{yz}$-orbitals 
in accordance with band-structure calculations\cite{liebsh}. The 
electronic properties of this model in application to Sr$_2$RuO$_4$ were 
studied recently and as was found can explain some features 
of the spin excitation spectrum in Sr$_2$RuO$_4$\cite{mazin,ng,morr,eremin}. 
However, this model fails to explain the observed magnetic anisotropy at low 
temperatures\cite{ishida} and line nodes in the superconducting 
order parameter below T$_c$ which are between the RuO$_2$-planes. 
On the other hand, it is known 
that the spin-orbit coupling plays an important role 
in the superconducting state of in Sr$_2$RuO$_4$\cite{ng}. 
This is further confirmed by the recent observation of the 
large spin-orbit coupling in the insulating Ca$_2$RuO$_4$\cite{saw}. 
Therefore, we include in our model spin-orbit 
coupling:
\begin{equation}
H_{so} = \lambda \sum_i {\bf L}_i {\bf S}_i
\quad,
\label{spinorbit}
\end{equation}
where the angular momentum {\bf L}$_i$ operates on the three 
$t_{2g}$-orbitals on the site $i$. Similar to an earlier approach\cite{ng}, we 
restrict ourselves to the three orbitals, ignoring $e_{2g}$-orbitals and 
choose the coupling constant $\lambda$ such that the t$_{2g}$-states behave 
like an $l=1$ angular momentum representation. Moreover, 
it is known that the quasi-two-dimensional $xy$-band is separated from 
the quasi-one-dimensional $xz$- and $yz$-bands. Then, one expects that the 
effect of spin-orbit coupling is small and can be excluded for 
simplicity. Therefore, we consider the effect of the spin-orbit coupling on 
$xz$- and $yz$-bands only. Then, the kinetic part of the
Hamiltonian $H_t +H_{so}$ can be diagonalized and the new energy 
dispersions are
\begin{eqnarray}
\epsilon_{{\bf k},yz}^{\sigma}  & = &  
(t_{{\bf k},yz}+t_{{\bf k},xz}  + A_{\bf k})/2  
\nonumber \\
\epsilon_{{\bf k},xz}^{\sigma}  & = &  
(t_{{\bf k},yz}+t_{{\bf k},xz}  - A_{\bf k})/2  
\label{disper}
\end{eqnarray}
where 
$A_{\bf k} = \sqrt{(t_{{\bf k},yz}-t_{{\bf k},xz})^2  + 
\lambda^2}$, and $\sigma$ refers to 
spin projection. 
\begin{figure}[t]
\centerline{\epsfig{clip=,file=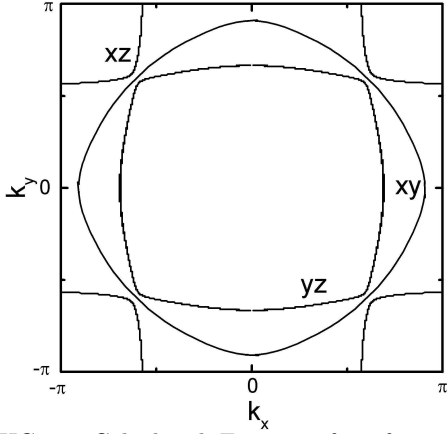,width=8.7cm,angle=0}}
\caption{Calculated Fermi surface for a RuO$_2$ plane in Sr$_2$RuO$_4$
taking into account spin-orbit coupling. 
}
\label{fig1}
\end{figure}
One clearly sees that the spin-orbit coupling does 
not remove the Kramers degeneracy of the spins. Therefore, the resultant 
Fermi surface is consists of three sheets like observed in the 
experiment. Most importantly, spin-orbit coupling together 
with Eq. (1) leads to a new quasiparticle which we label by pseudo-spin 
and pseudo-orbital indices. The unitary transformation 
$\tilde{U}_{\bf k}$ connecting 
old and new quasiparticles is defined for each wave vector and 
lead to the following relation between them:
\begin{eqnarray}
c^+ _{{\bf k},yz+}  & = &  u_{1{\bf k}} a^+ _{{\bf k},yz+} - i  
v_{1{\bf k}} a^+ _{{\bf k},xz+}, 
\nonumber\\
c^+ _{{\bf k},xz+}  & = &  u_{2{\bf k}} a^+ _{{\bf k},yz+} - i  
v_{2{\bf k}} a^+ _{{\bf k},xz+}, 
\nonumber\\
c^+ _{{\bf k},yz-}  & = &  u_{1{\bf k}} a^+ _{{\bf k},yz-} + i  
v_{1{\bf k}} a^+ _{{\bf k},xz-},
\nonumber\\
c^+ _{{\bf k},xz-}  & = &  u_{2{\bf k}} a^+ _{{\bf k},yz-} + i  
v_{2{\bf k}} a^+ _{{\bf k},xz-},
\label{tarnsf}
\end{eqnarray}
where $u_{m{\bf k}} = \frac{\lambda}{\sqrt{(t_{{\bf k}, yz} - t_{{\bf k}, xz} 
\mp A_{\bf k})^2 + \lambda^2}}$ and $v_{m{\bf k}} = \frac{t_{{\bf k}, yz} - 
t_{{\bf k}, xz} \mp A_{\bf k}}{\sqrt{(t_{{\bf k}, yz} - t_{{\bf k}, xz} 
\mp A_{\bf k})^2 + \lambda^2}}$. The '-' and '+' signs refer to the $m=1$ 
and $m=2$, respectively.  

In Fig.1 we show the resultant Fermi surfaces for each obtained band where  
we have chosen $\lambda = 100$meV in agreement with earlier 
estimations\cite{ng,saw}. One immediately sees that $xz$- and $yz$-bands 
split around the nested parts in good agreement with experiment\cite{shen}. 
Thus, spin-orbit coupling 
acts as a hybridization between these bands. However, in contrast to 
hybridization spin-orbit coupling introduces also an anisotropy 
for the states with pseudo-spins $\uparrow$ and $\downarrow$. This will  
be reflected in the magnetic susceptibility. Since the spin and orbital 
degrees of freedom are now mixed in some spin-orbital liquid, the magnetic 
susceptibility involves also the orbital magnetism which is very anisotropic.
\begin{figure}[t]
\centerline{\epsfig{clip=,file=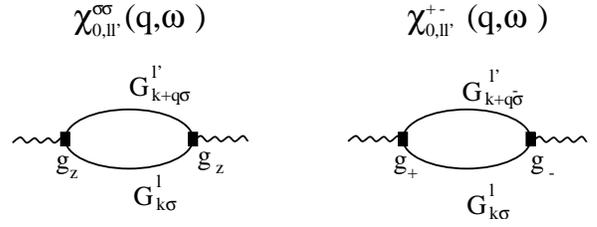,width=7.7cm,angle=0}}
\vspace{1ex}
\caption{ Diagrammatic representation for the transverse and longitudinal 
components of the magnetic susceptibility. The full lines represent 
the electron Green's function with pseudospin $\sigma$ and pseudo-orbital 
$l$-indexes. $g_{+}$ and $g_{z}$ denote the vertexes as described in the 
text. 
}
\label{fig2}
\end{figure}

For the calculation of the transverse, $\chi_l^{+-}$, and 
longitudinal, $\chi_l^{zz}$, components of the spin susceptibility of each 
band $l$ we use the diagrammatic representation shown in Fig. 2. Since 
the Kramers degeneracy is not removed by the spin-orbit coupling, the main 
anisotropy arises from the calculations of the anisotropic vertex
$g_z = \tilde{l}_z +2s_z$ and $g_+ = \tilde{l}_+ +2s_+$ calculated on the
basis of the new quasiparticle states. In addition, due to the hybridization 
between $xz$- and $yz$-bands we also calculate the transverse 
and longitudinal components of the the interband susceptibility 
$\chi_{ll'}$.  Then, for example,
\begin{eqnarray}
\chi_{0,xz}^{+-} ({\bf q}, \omega) =  - \frac{4}{N} \sum_{\bf k} &&
 (u_{2{\bf k}}u_{2{\bf k+q}}-v_{2{\bf k}}v_{2{\bf k+q}})^2 \times \nonumber\\
&& \frac{f(\epsilon_{{\bf k}xz}^{+})-f(\epsilon_{{\bf k+q}xz}^{-})}
{\epsilon_{{\bf k}xz}^{+} - \epsilon_{{\bf k+q}xz}^{-} +\omega +iO^+},
\label{lindpm}
\end{eqnarray} 
and 
\begin{eqnarray}
\lefteqn{\chi_{0,xz}^{zz} ({\bf q}, \omega) =   \chi_{xz}^{\uparrow} 
({\bf q}, \omega) +  \chi_{xz}^{\downarrow} ({\bf q}, \omega) = - \frac {2}{N} 
\sum_{\bf k}} && \nonumber\\
&& \left[u_{2{\bf k}} u_{2{\bf k+q}} + 
v_{2{\bf k}} v_{2{\bf k+q}} + \sqrt{2}( u_{2{\bf k}} v_{2{\bf k+q}} + v_{2{\bf k}} 
u_{2{\bf k+q}})\right]^2 \times \nonumber \\
&& \frac{f(\epsilon_{{\bf k}xz}^{+})-
f(\epsilon_{{\bf k+q}xz}^{+})}
{\epsilon_{{\bf k}xz}^{+} - \epsilon_{{\bf k+q}xz}^{+} +\omega +iO^+} 
\quad,
\label{lindzz}
\end{eqnarray}
where $f(x)$ is the Fermi function and $u_{\bf k}^2$ and $v_{\bf k}^2$ 
are the corresponding coherence factors which we have  
\begin{figure}
\centerline{\epsfig{clip=,file=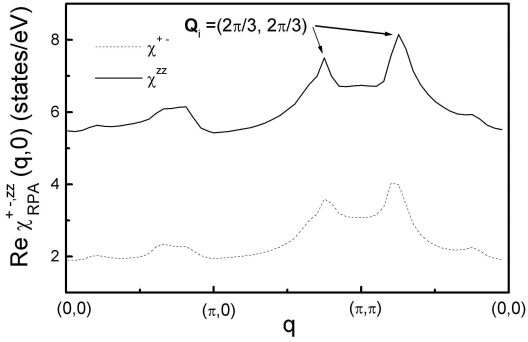,width=7.7cm,angle=0}}
\caption{Results for the real part of the out-of-plane (solid curve) 
 and in-plane (dashed curve) magnetic susceptibilities,
Re $\chi({\bf q},\omega)$, calculated within RPA using $U = 0.575$eV 
along the route 
$(0,0)\rightarrow (\pi,0) \rightarrow (\pi,\pi) \rightarrow (0,0)$ 
within the first Brillouin Zone at temperature T = 100K.  
}
\label{fig3}
\end{figure}
calculating through the corresponding vertexes  using Eq. (4). 
For all other orbitals the calculations 
are straightforward. Note, that the magnetic response of the $xy$-band remains 
isotropic. 
\begin{figure}[t]
\centerline{\epsfig{clip=,file=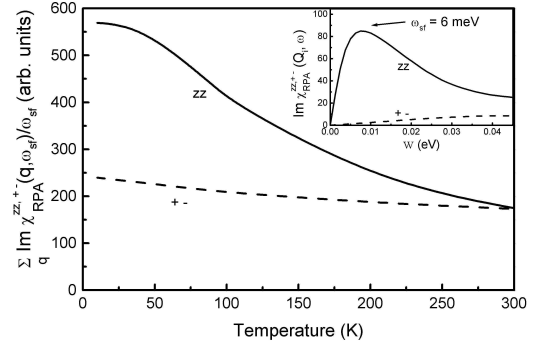,width=7.5cm,angle=0}}
\caption{ Temperature dependence of the imaginary part of the 
spin susceptibility divided by $\omega_{sf}$ and summed over {\bf q}. Note,  
$zz$ and $+-$ refer to the out-of-plane (solid curve) 
and in-plane (dashed curve)
components of the RPA spin susceptibility. In the inset we show the 
corresponding frequency dependence of the 
Im$\chi_{RPA} ({\bf Q_{i}}, \omega)$ at the IAF wave vector 
${\bf Q}_i = (2\pi/3, 2\pi/3)$. The results 
for the out-plane component (solid curve) are in a quantitative 
agreement with INS experiments\protect\cite{sidis}. 
}
\label{fig4}
\end{figure}
One clearly sees the difference between longitudinal and transverse components 
which results from the calculated matrix elements. Moreover, the longitudinal 
gets an extra term due to $\tilde{l}_z$ while the transverse does not contain 
the contributions from $\tilde{l}_+$ or $\tilde{l}_-$. The latter occur 
due to the fact that $xz$- and $yz$-states are a combination of the real 
orbital states $|2,+1>$ and $|2,-1>$. Thus the transition between these 
two states are not possible with  $\tilde{l}_+$ or $\tilde{l}_-$ operators. 
Therefore, 
each component of the longitudinal susceptibility gets an extra term 
in the matrix element that sufficiently enhances their absolute values.

Assuming $U_{ij}=\delta_{ij}U$ one gets the following expressions
for the transverse susceptibility within RPA:
\begin{eqnarray}
\chi^{+-}_{RPA,l}({\bf q}, \omega) & = & 
\frac{\chi^{+-}_{0,l}({\bf q}, \omega)}{1-U\chi^{+-}_{0,l}({\bf q}, \omega)}
\quad,
\label{RPApm}
\end{eqnarray}
and for the longitudinal susceptibility 
\begin{eqnarray}
\lefteqn{\chi^{zz}_{RPA,l}({\bf q}, \omega) =} && \nonumber\\ 
&& \frac{\chi^{\uparrow}_{0,l}({\bf q}, \omega)+
\chi^{\downarrow}_{0,l}({\bf q}, \omega)+2U
\chi^{\uparrow}_{0,l}({\bf q}, \omega)
\chi^{\downarrow}_{0,l}({\bf q}, \omega)}{1-U^2
\chi^{\downarrow}_{0,l}({\bf q}, \omega)
\chi^{\uparrow}_{0,l}({\bf q}, \omega)} 
\quad.
\label{RPApzz}
\end{eqnarray}

In Fig. 3 we show the results for the real part 
of the transverse and longitudinal total susceptibility,
$\chi^{+-,zz}_{RPA} = \sum_{i} \chi^{+-,zz}_{RPA,i}$ along the route 
$(0,0)\rightarrow (\pi,0) \rightarrow (\pi,\pi) \rightarrow (0,0)$ 
in the first Brillouin Zone for  
$U=0.505$eV . Note, the important difference between the two components.
The longitudinal component of the spin susceptibility is almost three times 
larger than the transverse one all over the Brillouin Zone. Moreover, 
despite of some structure seen in $\chi^{+-}_{RPA}$ at $(2\pi/3, 2\pi/3)$
there are no real incommensurate antiferromagnetic 
fluctuations at this wave vector. On the other 
hand, the structure in $\chi^{zz}_{RPA}$ at the same wave vector refers 
to real fluctuations. The latter is seen in the inset of Fig.4 where 
we present the results for the
frequency dependence of the imaginary part of the total susceptibilities at 
{\bf Q}$_i = (2\pi/3, 2\pi/3)$ and temperature T=20K.
The longitudinal component reveals a peak at 
approximately $\omega_{sf} = 6$meV in quantitative agreement with 
experimental data on INS\cite{sidis}. The transverse component 
is featureless showing the absence of the incommensurate antiferromagnetic 
spin fluctuations. 
Thus, the fluctuations in the transverse susceptibility are isotropic 
and ferromagnetic-like. Therefore, antiferromagnetic fluctuations are 
present only perpendicular to the RuO$_2$-plane. 

We also note that our results are in accordance with earlier estimations 
made by Ng and Sigrist\cite{ng2} with one important difference. In their work 
it was found that the IAF are slightly enhanced in the longitudinal 
components of the $xz$- and $yz$-bands in comparison to the transverse one.
In our case we have found that the longitudinal 
component of the magnetic susceptibility strongly enhances due to 
otbital contributions. Moreover, we show by taking into account 
the correlation 
effects within random-phase-approximation(RPA) the IAF are further 
enhanced in the $z$-direction. 

In order to see the temperature dependence of the magnetic 
anisotropy induced by the spin-orbit coupling we display in Fig.4 the 
temperature dependence of the quantity  
$\sum_{\bf q} \frac{Im \chi_{RPA} ({\bf q}, \omega_{sf})}{\omega_{sf}}$ 
for both components. At room temperatures both longitudinal and transverse 
susceptibilities are almost identical, since thermal effects wash out the 
influence of the spin-orbit interaction. With decreasing temperature 
the magnetic anisotropy arises and at low temperatures we find 
the important result that the out-of-plane 
component $\chi^{zz}$ is about two times larger than the in-plane one 
($\chi^{zz}>\chi^{+-}/2$). 
\begin{figure}[t]
\centerline{\epsfig{clip=,file=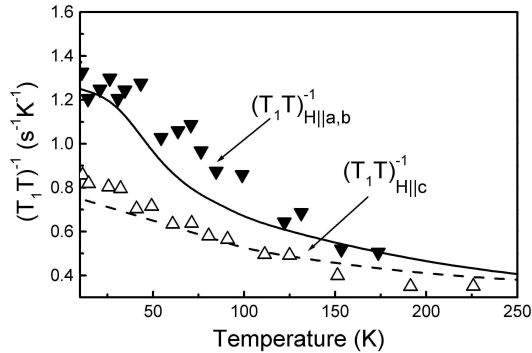,width=7.6cm,angle=0}}
\caption{ Calculated normal state temperature dependence of the 
nuclear spin-lattice relaxation rate $T_1^{-1}$ of $^{17}$O in the 
RuO$_2$-plane for the 
external magnetic field applied along $c$-axis (dashed curve) 
and along the $ab$-plane (solid curve). Down- and up-triangles are 
experimental points taken from Ref. \protect\onlinecite{ishida} 
for the corresponding magnetic field direction.  
}
\label{fig5}
\end{figure}

Finally, 
in order to compare our results with experimental data we calculate the 
nuclear spin-lattice relaxation rate for $^{17}$O ion in the RuO$_2$-plane 
for different external magnetic field orientation 
$(\text{i = }a,b, \text{ and   }c)$ 
\begin{eqnarray}
\left[\frac{1}{T_1 T}\right]_i & = & \frac{2k_B \gamma^2_n}{(\gamma_e \hbar)^2}
\sum_{\bf q} |A_{\bf q}^{p}|^2 \frac{\chi''_{p} ({\bf q}, \omega_{sf})}
{\omega_{sf}}
\quad,
\label{t1t}
\end{eqnarray}  
where $A_{\bf q}^{p}$ is the 
$q$-dependent hyperfine-coupling constant perpendicular to the 
$i$-direction. 

In Fig.5 we show the calculated temperature dependence 
of the spin-lattice relaxation for an external magnetic field 
within and perpendicular to the RuO$_2$-plane together with experimental data. 
At $T = 250$K the spin-lattice relaxation rate is almost isotropic. 
Due to the anisotropy in the 
spin susceptibilities arising from spin-orbit coupling 
the relaxation rates become different with decreasing temperature. 
The largest anisotropy occurs close to the 
superconducting transition temperature in 
good agreement with experimental data\cite{ishida}.

To summarize, our results clearly demonstrate the 
essential significance of spin-orbit 
coupling for the spin-dynamics already in the normal state of the 
triplet superconductor Sr$_2$RuO$_4$. We find that the 
magnetic response becomes strongly anisotropic even within a RuO$_2$-plane: 
while the in-plane response is mainly ferromagnetic, the out-of-plane 
response is antiferromagnetic-like.

Let us also remark on the implication of our results for the 
triplet superconductivity in Sr$_2$RuO$_4$. In a previous study\cite{eremin}, 
neglecting spin-orbit coupling but including 
the hybridization between $xy$- and $xz-, yz$-bands, 
we have found ferromagnetic and 
IAF fluctuations within the $ab$-plane.This would lead to nodes 
within the RuO$_2$-plane. However, due to the magnetic anisotropy 
induced by spin-orbit coupling, a nodeless $p-$wave pairing is possible 
in the RuO$_2$-plane as experimentally observed. Our results provide 
further evidence for the importance of 
spin fluctuations for triplet superconductivity in Sr$_2$RuO$_4$.

We are thankful for stimulating discussions with B.L. Gyorffy, Y. Maeno, 
D. Fay, M. Eremin, R. Tarento and M. Ovchinnikova for critical reading 
of the manuscript. We are grateful to 
German-French Foundation (PROCOPE) for the financial support. 
The work of I. E. is supported by the Alexander von Humboldt Foundation 
and CRDF Grant No. REC. 007.  
\end{document}